# Multi-phonon-assisted absorption and emission in semiconductors and its potential for laser refrigeration

Jacob B. Khurgin

Johns Hopkins University, Baltimore MD 21208

jakek@jhu.edu

Laser cooling of semiconductors has been an elusive goal for many years, and while attempts to cool the narrow gap semiconductors such as GaAs are yet to succeed, recently, net cooling has been attained in a wider gap CdS. This raises the question of whether wider gap semiconductors with higher phonon energies and stronger electron-phonon coupling are better suitable for laser cooling. In this work we develop a straightforward theory of phonon-assisted absorption and photoluminescence of semiconductors that involves more than one phonon and use to examine wide gap materials, such as GaN and CdS and compare them with GaAs. The results indicate that while strong electron-phonon coupling in both GaN and CdS definitely improves the prospects of laser cooling, large phonon energy in GaN may be a limitation, which makes CdS a better prospect for laser cooling.

Optical refrigeration of solids [1-4], which traces its origins to Pringsheim's pioneering proposal made in 1929 [5] offers an attractive alternative to more traditional means of cooling as, on one hand, unlike most cryogenic coolers it has no moving parts, and on the other hand, its efficiency can potentially be higher than in thermoelectric coolers. The mechanism of laser refrigeration (LR) of solids is an anti-Stokes photoluminescence (PL), a process in which the optical absorption is either directly involves or is quickly followed by the absorption of phonons thus removing the heat from the lattice. Since the concept of LR of solids has been first experimentally verified in 1995 [6], LR all the way to cryogenic temperatures of 119 has been achieved [7,8] in $Yb^{3+}$-doped $LiYF_4$ crystals bringing LR close to practical applications. But it would be far more preferable to use semiconductors rather than rear earth doped crystals and ceramics as laser-cooled heat sinks because then the heat sinks could also serve as substrates for the electronic devices, such as IR detectors, whose heat they are supposed to remove, thus enabling integration. The conditions for LR of semiconductors have been formulated a decade ago [9], and in the ensuing years an effort was mounted to cool the most widely available and well studied direct bandgap semiconductor –GaAs. Significant progress has been made in improving the quality of GaAs samples and reducing all the processes that impede LR – defects, surface recombination, parasitic absorption and other [10], and while the efficiency of AS PL approached 99%, no net LR has been achieved yet, even though the latest results do come tantalizingly close.

Then, in 2013, it was first reported that net LR has been achieved in a rather unexpected II-VI semiconductor medium – cadmium sulfide [11,12] with a wider bandgap and higher polarity than GaAs. This groundbreaking discovery has immediately made the case for expanding the scope of the search for the material for LR even further expanded, as was indeed done in [13] where strong AS PL has been observed in wide bandgap GaN. The arguments being made for using wide bandgap materials are that the Froehlich interaction between the electrons and phonons are stronger in wide gap materials and, in case of GaN phonon energies themselves are higher, as evidenced from numerous Raman studies. But other than these "common sense" arguments no rigorous evaluation of LR potential of various semiconductors has been made, other than early works [14-16] in which it was assumed that absorption occurred into band-edge of exciton states and the Auger recombination and absorption saturation were limiting factors. Since then, it had become obvious that the best results were always obtained in the highest purity samples while the laser was always tuned substantially below the nominal bandedge energy of the material (by as much as 70meV in [11] and 170meV in [13]), eliminating the possibility that cooling occurs via impurity and defect associated defects states or excitons and giving a strong indication that phonons must be involved into the optical processes at every step, and not just serve to thermalize the photo-excited carriers.

T     he phonon-assisted absorption is semiconductors has been studied since 1950's [17] when exponentially decaying bandtail absorption has been first observed and given a name "Urbach tail". Numerous experimental and theoretical studies have been performed (see review in [18]) and the exponential shape of the bandtail has been explained, even though the widely used approximate expression [17] for the Urbach tail absorption: $\alpha \sim \alpha_0 \exp[\sigma(\hbar\omega_0 - \hbar\omega)/kT]$ suffers from the fact that the steepness parameter $\sigma$ does not directly relate to the strength of electron-phonon interaction, and the expression does not merge seamlessly into the expression for the inter-band absorption proportional to $\sqrt{\omega - \omega_{gap}}$. Furthermore, all the work on Urbach tail was historically performed at low light intensity, and often at low temperatures, hence the exciton effects of were overemphasized over the band-to-band absorption, and, the two effects that are of greatest importance to LR , Burstein-Moss (BM) shift of phonon-assisted transitions, and phonon-assisted PL have not been given proper attention. In our prior work [19] we have considered phonon-assisted processes and their effect on GaAs LR but only looked at single-phonon-assisted transitions, and used simplified relations between the absorption and PL which did not account for the possibility of phonon-assisted optical gain. The experiments [11-13], demonstrate that the large AS PL shifts are achieved when the detuning from band edge greatly exceeded the phonon energy indicating that multi-phonon processes may be the key to understanding the LR in materials with strong electron-phonon coupling.

In this work we develop a simple and rigorous theory of multi-phonon assisted absorption and emission in the polar semiconductors and apply this theory to three different semiconductors, GaAs, CdS, and GaN ,in order to ascertain their relative merits for LR.. The theory differs from the earlier works [20,21] in that it considers all relevant processes in conduction (CB) and valence (VB) bands, resulting in seemless transition between the regions of 2-phonon, 1-phonon and direct absorption, and takes into account photo-generated carriers in the bands leading to effective B-M shift [22]. The diagram of the relevant processes for the two-phonon assisted absorption (Fig.1) shows four different pathways leading to generation of electron-hole pair with wave vectors separated by energy $\hbar\omega + 2\hbar\Omega$ where $\omega$ is a photon and $\Omega$ is a LO phonon frequency. The first pathway, **CC** , consists of direct non-resonant optical transition from $k_v$ to $k_c$ followed by two consecutive phonon absorption processes in the CB –first to intermediary state $k_c'$ and then to final state $k_c''$  path **VV** direct transition into $k_c$ is followed by two phonon absorption processes in the VB. In pathway **CV** the photon absorption is followed first by phonon absorption in the CB and then in VB, while in path **VC** the order of phonon absorption is reversed. The absorption coefficient of the photon of frequency $\omega$ can be found by integration over the pairs of states $k_c, k_v$ involved in the photon transition, as roughly

$$\alpha(\omega) = \frac{2}{3n}\alpha_0 \left(\frac{2\mu}{\hbar^2}\right)^{1/2} \int_0^\infty L_{cv,k}(\hbar\omega) E_{cv,k}^{1/2} dE_{cv,k} \tag{1}$$

where $\alpha_0 = 1/137$ is a fine structure constant, $\mu$ is the reduced mass, $E_{cv,k} = \hbar^2 k^2 / 2\mu$ and the Lorentzian function under the integral is

$$L_{cv,k} = \frac{1}{\pi} \frac{(\Gamma_c + \Gamma_v)/2}{(E_{cv,k} + E_g - \hbar\omega)^2 + (\Gamma_c + \Gamma_v)^2/4} \tag{2}$$

The terms in the numerator.

$$\Gamma_{c(v)} = \frac{1}{2\pi} \frac{e^2 \hbar \Omega n_{LO}}{2\varepsilon'} \frac{2m_{c(v)}}{\hbar^2} \frac{1}{k} \int_0^\infty \ln\left|\frac{k + k'_{c(v)}}{k - k'_{c(v)}}\right| L'_{c(v)} dE'_{c(v)} \tag{3}$$

Here represent the two-phonon scattering strength in which the first scattering occurs always in CB (VB), the integration is over all the intermediate states $k'_{c(v)}$ in CB (VB), the phonon occupation number is $n_{LO} = 1/[\exp(\hbar\omega_{LO}/kT) - 1]$, $(\varepsilon')^{-1} = \varepsilon_0^{-1} - \varepsilon_\infty^{-1}$, and the new Lorentzian is

$$L'_{c(v)} = \frac{1}{\pi} \frac{\left(\Gamma'_{cc(vv)} F^a_{cc(vv)} + \Gamma'_{cv(vc)} F^e_{cv(vc)}\right)/2}{\left(E'_{c(v)} + E_{v(c)} + E_g - \hbar\omega - \hbar\Omega\right)^2 + \left(\Gamma'_{cc(vv)} + \Gamma'_{cv(vc)}\right)/2} \tag{4}$$

The new scattering terms in (4) describe the strength of one-phonon scattering into the final states for **CC** and **VV** processes,

$$\Gamma_{cc(vv)} = \frac{1}{2\pi} \frac{e^2 \hbar \Omega n_{LO}}{2\varepsilon'} \frac{1}{k'_{c(v)}} \ln\left|\frac{k'_{c(v)} + k''_{c(v)}}{k'_{c(v)} - k''_{c(v)}}\right| \frac{2m_{c(v)}}{\hbar^2}, \tag{5}$$

and these terms are multiplied in (4) by the absorption Fermi factors $F^a_{cc(vv)} = [1 - f_{v(c)}(k_{v(c)})][1 - f_{c(v)}(k''_{c(v)})]$ where $f_v(k_v)$ is a probability of having a hole in the VB. For **CV** and **VC** processes, expressions similar to (5) are easily obtained. For the phonon assisted gain caused by stimulated emission of the photons one can still use (1) but with $n_{LO} + 1$ in place of $n_{LO}$ in (3) and (5), and the emission Fermi factor $F^e_{cc(vv)} = f_{v(c)}(k_{v(c)}) f_{c(v)}(k''_{c(v)})$ in (4). The net absorption that includes BM shift via state filling and PL can then be obtained. The salient feature of our approach is that when detuning from band edge is less than one phonon energy the Lorentzian (4) behaves like a delta function and a single phonon assisted transition is obtained, and, when phonon energy is larger than bandgap, the Lorentzian (2) becomes a delta function leading to a normal parabolic band edge. Although we have considered only two phonon assisted processes, the theory can be easily expanded to three or more phonons, but our calculations (confirmed by experiments [11-13]) indicate that absorption below two phonon energies from the bandgap is weak and saturates too easy for LR.

Let us now turn our attention to the results, where in Fig.2a. we first look at the absorption spectra of GaN with a very low carrier density at different temperatures from 100K to 350K with 50K interval. As one can see the Urbach tail extends all the way to $2\hbar\Omega \sim 180 meV$ below the direct absorption edge. At low temperatures one can clearly see the steps associated with the transition from two phonon-assisted to one-phonon assisted to direct absorption, but at higher temperatures the bandtail loses its features and becomes exponential. At elevated temperature is merges smoothly with the direct absorption. Near the direct band edge at all temperatures the slope agrees well with the experimental data [23]. Next, at Fig.2a one can see the net absorption spectra of GaN with photo-excited carrier density equal to N=$10^{18}$cm$^{-3}$. The BM shift is evident, especially at low temperatures. The shift is very large, because for the transition assisted by the m phonons maximum gain and maximum absorption at a given frequency are related as $g_{max}(\omega)/\alpha_{max}(\omega) = (n_{LO}+1)^m/n_{LO}^m = \exp(m\hbar\Omega/kT)$ - hence due to large (~91 meV) phonon energy in GaN the absorption in the Urbach tail quickly saturates and turns into gain, particularly at lower temperatures. When compared with PL spectra in Fig. 2c one can see that at low temperatures AS shift is barely present – there is a very strong Stokes-shifted phonon assisted PL coming from Urbach tail, and with absorption in the tail completely inverted no net cooling can take place. Therefore, it appears that large phonon energy may be a handicap rather than advantage when it comes to LR .

Next we compare the bandtail absorption and PL of three different semiconductors at T=300K for different carrier concentrations. The first one, GaAs is a medium bandgap ( $E_g \sim 1.5 eV$ )semiconductor with a medium LO phonon energy ( $\hbar\Omega = 36 meV$ ) and relatively weak Froehlich interaction ( $1/\varepsilon' = 0.014$ ). As one can see in Fig.3.a bandtail absorption is rather weak, but at the same time PL, shown in Fig.3b is also quite weak below bandgap hence the AS shift at low carrier densities is substantial, about 70 meV. In Fig.3c we have plotted the BM shift of the absorption edge (defined here as energy at which net absorption equals 10 cm$^{-1}$) and the mean PL energy. The BM shift becomes significant at carrier densities of roughly $10^{17}$ cm$^{-3}$ and at higher densities of $10^{18}$ cm$^{-3}$ the AS shift of PL disappears completely Since radiative recombination rate at low densities is small, the impact of nonradiative recombination becomes makes net cooling difficult, and at higher pumping densities the cooling power decreases as the energy difference between absorption edge and PL becomes less and less.

If one now turns attention to Fig. 4, where results for GaN – material with a wide bandgap ($E_g \sim 3.4 eV$), very large LO phonon energy ($\hbar\Omega = 88 meV$) and very strong Froehlich interaction ($1/\varepsilon' = 0.077$) are plotted. A strong bandtail absorption (Fig.4a) at low-to-moderate densities of photo-excited carriers keeps absorption edge well below the bandgap, but Stokes shifted phonon assisted PL is also strong (Fig. 4a) and the peak of PL stays below the band edge. Hence the total PL AS shift is only marginally stronger than in GaAs. Moreover, just as in GaAs the AS shift quickly diminishes at increased carrier densities as phonon assisted gain overwhelms the absorption. As mentioned above, this is a rather unappreciated fact. For the direct absorption, the maximum gain at a given frequency in pumped semiconductor asymptotically approaches the absorption at that frequency in the unpumped semiconductor, but when absorption is phonon assisted, the maximum gain can be orders of magnitude higher than that, especially if the phonon energy is high. This can have favorable consequences for the lasers but for optical refrigeration it leads to large Burstein-Moss shift and reduction in cooling power.

This brings us to Fig.5 and the third semiconductor, CdS, characterized by a relatively wide band gap ($E_g \sim 2.42 eV$), moderate phonon energy ($\hbar\Omega = 38 meV$) and strong Froehlich interaction ($1/\varepsilon' = 0.044$). As shown in Fig.5a the Urbach tail does not extend as deep as in GaN, but, at the same time PL (Fig.5b) is significantly above than band edge – hence the total AS shift of PL (Fig.3A) is on the order of 100meV – just as large as in GaN. At higher carrier concentrations the bane edge does not shift upward as fast as in GaN because the ratio between maximum gain and absorption $(n_{LO}+1)^m / n_{LO}^m$ at room temperature is two orders of magnitude less for two-phonon assisted transitions in CdS than in GaN due to lower phonon energy. As a result, a decent (~55meV) AS shift can be maintained even at densities of $2.5 \times 10^{18} cm^{-3}$ - this density is about 35% higher than in GaN, and, since the photon energy in CdS is 30% less than in GaN it indicates that 80% higher cooling power efficiency can be attained in CdS.

To conclude, we have developed a theory for multi-phonon assisted absorption, gain, and PL in direct bandgap semiconductors and applied to it GaAs, GaN and CdS. In all three cases large AS shifts between absorption edge and PL are attainable, but the cooling power is limited by the B-M shift which reduces the AS shift at higher carrier densities. These densities are well below the onset of Auger recombination, hence we confirm observation [19] that state filling (B-M shift) is the main fundamental limitation of optical refrigeration of semiconductors. Of course, presently LR is still limited by parasitic effects of nonradiative recombination and background absorption, but it still important to know the fundamental constraints.

When it comes to comparison between three materials relatively weak Froehlich interaction and small effective mass of GaAs make Urbach tail absorption weak and saturable at relatively low carrier densities which is of course makes achieving LR in it more difficult, although possible in good quality material. Stronger Froehlich interaction in more polar CdS and GaN makes them both more favorable candidates for laser refrigeration than GaAs, but, perhaps in a bit of surprise, large phonon energy in GaN puts it at relative disadvantage since so few ($n_{LO}$~0.03) phonons are excited even at room temperature. CdS, on the other hand, has relatively strong Froehlich interaction and quite a few ($n_{LO}$~0.3) phonons are available for cooling at room temperature, which is, overall, a winning combination. While this does not necessarily mean that CdS and similar polar materials are absolutely the best choice, one can say that the fact that optical refrigeration in a semiconductor has been first attained in CdS [11] is not a coincidence.

The author acknowledges support of NSF DMR-1207245

**Figure captions**

**Figure 1**. Diagrams of four pathways involved in the two-phonon assisted absorption in direct bandgap semiconductors

**Figure 2.** **(a)** absorption of unexcited GaN at different temperatures between 100K and 350K **(b)** Net absorption, and **(c)** PL of GaN with concentration of photo-excited carriers equal to $10^{18}$ cm$^{-3}$

**Figure 3** **(a)** Net absorption, and **(b)** PL of GaAs at T=300K for different concentrations of photo-excited carriers **(c)** – mean PL and Burstein Moss shift of absorption edge with carrier concentration at 300K

**Figure 4** **(a)** Net absorption, and **(b)** PL of GaN at T=300K for different concentrations of photo-excited carriers **(c)** – mean PL and Burstein Moss shift of absorption edge with carrier concentration at 300K

**Figure 5** **(a)** Net absorption, and **(b)** PL of CdS at T=300K for different concentrations of photo-excited carriers **(c)** – mean PL and Burstein Moss shift of absorption edge with carrier concentration at 300K

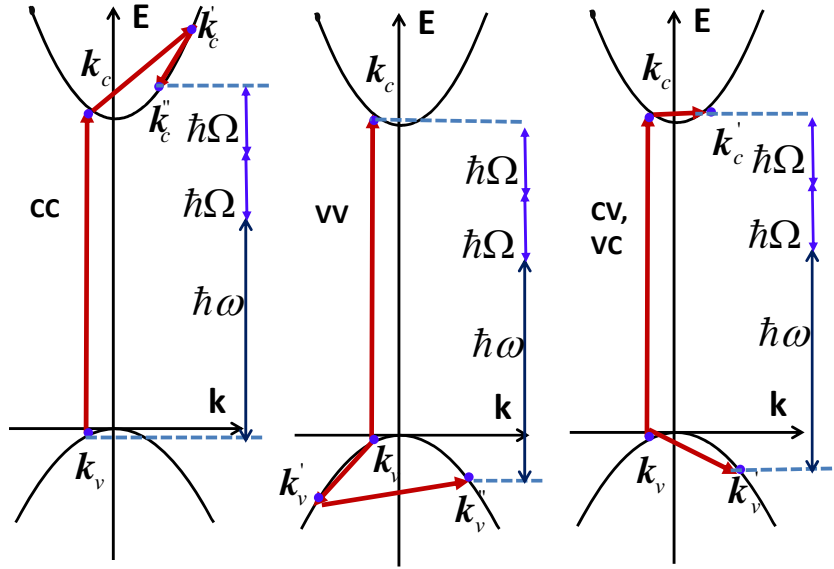

**Figure 1**. Diagrams of four pathways involved in the two-phonon assisted absorption in direct bandgap semiconductors

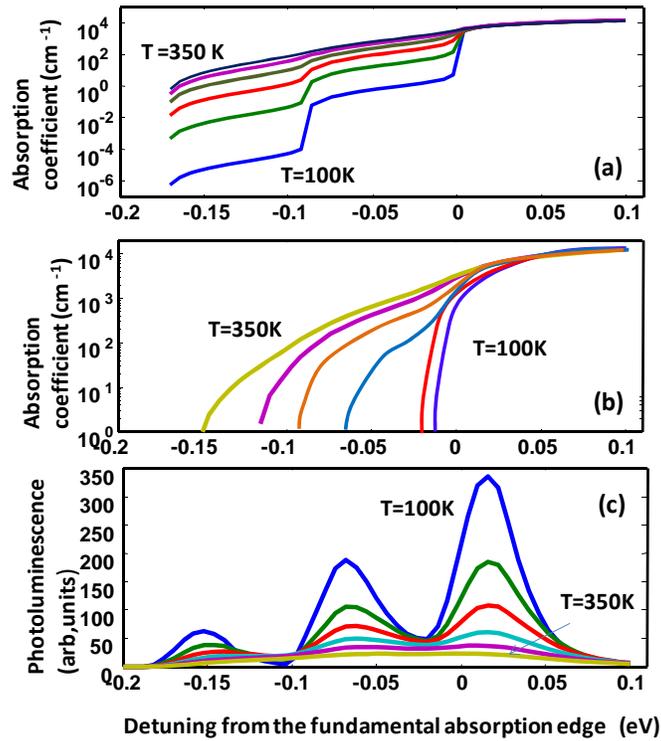

**Figure 2.** (**a**) absorption of unexcited GaN at different temperatures between 100K and 350K (**b**) Net absorption, and (**c**) PL of GaN with concentration of photo-excited carriers equal to $10^{18}$ cm$^{-3}$

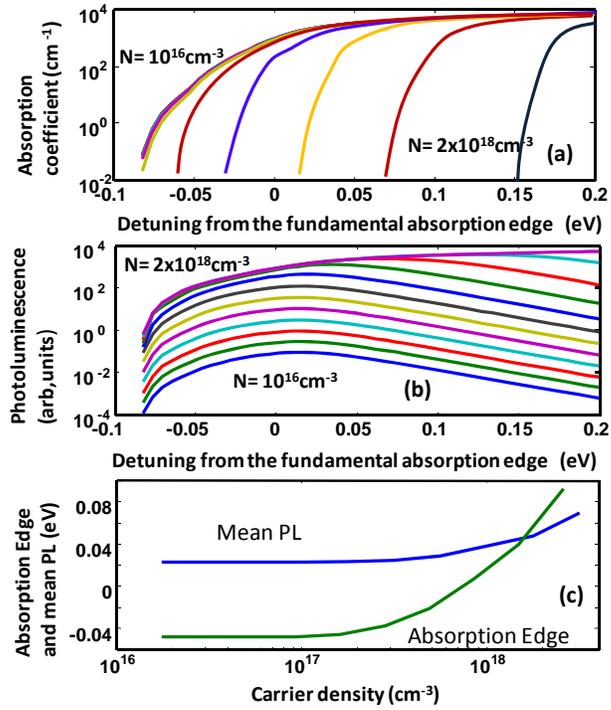

**Figure 3 (a)** Net absorption, and **(b)** PL of GaAs at T=300K for different concentrations of photo-excited carriers **(c)** – mean PL and Burstein Moss shift of absorption edge with carrier concentration at 300K

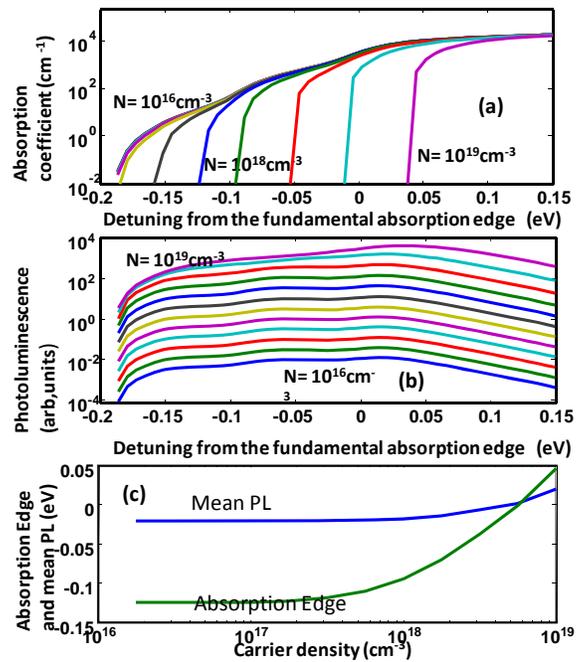

**Figure 4 (a)** Net absorption, and **(b)** PL of GaN at T=300K for different concentrations of photo-excited carriers **(c)** – mean PL and Burstein Moss shift of absorption edge with carrier concentration at 300K

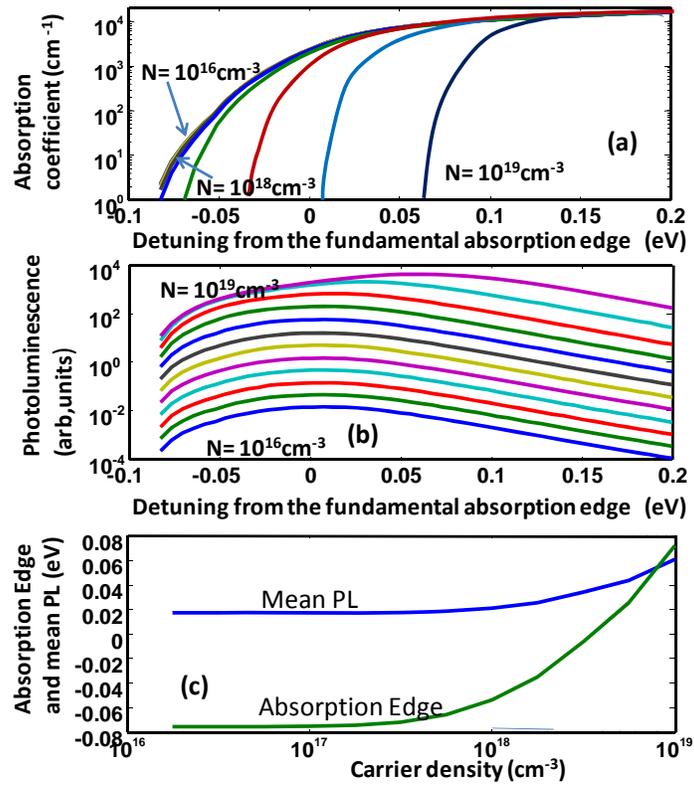

**Figure 5 (a)** Net absorption, and **(b)** PL of CdS at T=300K for different concentrations of photo-excited carriers **(c)** – mean PL and Burstein Moss shift of absorption edge with carrier concentration at 300K